# Dynamics and Structure of Monolayer Polymer Crystallites on Graphene

*Max Gulde[1,†,*], Anastassia N. Rissanou[2], Vagelis Harmandaris[2,3,*], Marcus Müller[4], Sascha Schäfer[1], Claus Ropers[1]*

[1]4th Physical Institute - Solids and Nanostructures, University of Göttingen, 37077, Göttingen, Germany
[2]Institute of Applied and Computational Mathematics, Foundation for Research and Technology Hellas, 71110 Heraklion, Crete, Greece
[3]Department of Mathematics and Applied Mathematics, University of Crete, 71409, Heraklion, Crete, Greece
[4]Institute for Theoretical Physics, University of Göttingen, 37077, Göttingen, Germany

*\* Corresponding authors: max.gulde@gmail.com, harman@uoc.gr*

ABSTRACT
Graphene-based nanostructured systems and van-der-Waals heterostructures comprise a material class of growing technological and scientific importance. Joining materials with vastly different properties, polymer-graphene heterosystems promise diverse applications in surface- and nanotechnology, including photovoltaics or nanotribology. Fundamentally, molecular adsorbates are prototypical systems to study confinement-induced phase transitions exhibiting intricate dynamics, which require a comprehensive understanding of the dynamical and static properties on molecular time and length scales. Here, we investigate the dynamics and the structure of a single polyethylene chain on free-standing graphene by means of molecular dynamics simulations. In equilibrium, the adsorbed polymer is orientationally linked to the graphene as two-dimensional folded-chain crystallites or, at elevated temperatures, as a floating solid. The associated superstructure can be reversibly melted on a picosecond time scale upon quasi-instantaneous substrate heating, involving ultrafast heterogeneous melting *via* a transient floating phase. Our findings elucidate time-resolved molecular-scale ordering and disordering phenomena in individual polymers interacting with solids, yielding complementary information to collective friction and viscosity, and linking to recent experimental observables from ultrafast electron diffraction. We anticipate that the approach will help in resolving non-equilibrium phenomena of hybrid polymeric systems over a broad range of time and length scales.

KEYWORDS
polymer, graphene, dynamics, crystallite, floating solid, two-dimensional phase

MAIN TEXT

Polymeric-based nanostructured materials encompass a wide class of chain-like molecules with very diverse thermal, electronic, rheological and mechanical characteristics.[1-3] These properties are not only a result of the chemical structure of the individual monomers, but are also strongly influenced by the molecular structure, including chain length, relative alignment or crystallinity, as well as interchain interactions.[4-5] Therefore, it is expected that spatially confined polymers, *e.g.*, as atomically thin films on surfaces or in polymer nanocomposites, will exhibit behavior strongly differing from the bulk.[6-8] Indeed, various properties of polymers in the proximity of polymer/solid interfaces and interphases differ from their bulk values.[9-13] The macroscopic behavior of such hybrid systems follows from a subtle interplay between conformational entropy of polymer chains and the adhesive interaction between polymer chains and a corrugated substrate. This interplay gives rise to alignment and registration of molecules, may even induce a liquid-solid transition, and results in a complex dynamical behavior of chains at the interface.[14-19] However, the atomic-level study of such systems, at equilibrium and especially under non-equilibrium conditions, presents a considerable challenge, both for experimental approaches and for theoretical modeling and simulations.

From the experimental point of view, investigations of the structural evolution of ultrathin polymer layers on their characteristic length and time scales requires non-invasive methods with monolayer sensitivity and atomic scale resolution. Recent advances in the field of time-resolved electron diffraction allowed for observations of the melting of a polymer crystal of monolayer thickness on graphene, giving first insights into the dynamics of such a system.[20] Theoretical studies have mainly focused on the heat transfer from a solid substrate or nanoparticle to a fluid macromolecular phase,[21-23] and, additionally, on the formation and melting of a 2D free-standing polymer crystal supported by a thin surface.[24] Computer simulations have also provided insights into the crystallization of polymers on substrates[25, 26] and the melting of single-chain crystallites in the bulk.[27] Such studies raise many fundamental questions about the two-dimensional melting process on the atomic level: What are the relevant time and length scales for the structure and the dynamics of a single polymer chain? What is the influence of the substrate, *e.g.*, on the molecular conformation of the adsorbate? What is the nature of the superstructure melting and the corresponding thermal transport, particularly in a situation far away from thermal equilibrium? How is this process affected by the quasi-two-dimensionality of such systems? How does the re-crystallization take place?

Molecular dynamics (MD) simulations present a valuable tool to answer such questions, providing us with direct quantitative means to study the trajectories of single atoms during complex processes and transitions for specific polymer/substrate nanostructured systems. MD is a method that was used in the past to reveal structural and dynamical aspects of various amorphous polymer films, either supported by graphene or confined between graphene sheets at equilibrium.[28,15,12,13] In this work, we employ all-atom MD simulations to investigate the structural and dynamical aspects of a single polyethylene (PE) chain adsorbed on free-standing graphene. More specifically, we predict the characteristic time scales associated with structural changes during the formation and the melting of a polymer crystal. In addition, we also provide insight into the physical mechanisms of the superstructure melting process, as well as into the energy transfer from the graphene to the adsorbed polymer. Such interfaces are of particular importance due to the exceptional properties of graphene (*e.g.*, electron transport capacity, high intrinsic tensile strength and stiffness, high thermal and electrical conductivity, impermeability for all gas molecules and transparency, *etc.*), which make this material a promising candidate for an enormous range of possible technological applications, such as the reinforcement of polymer nanocomposites.[29,30,9,10]

**Monolayer Folded-Chain Polymer Crystals in Equilibrium**

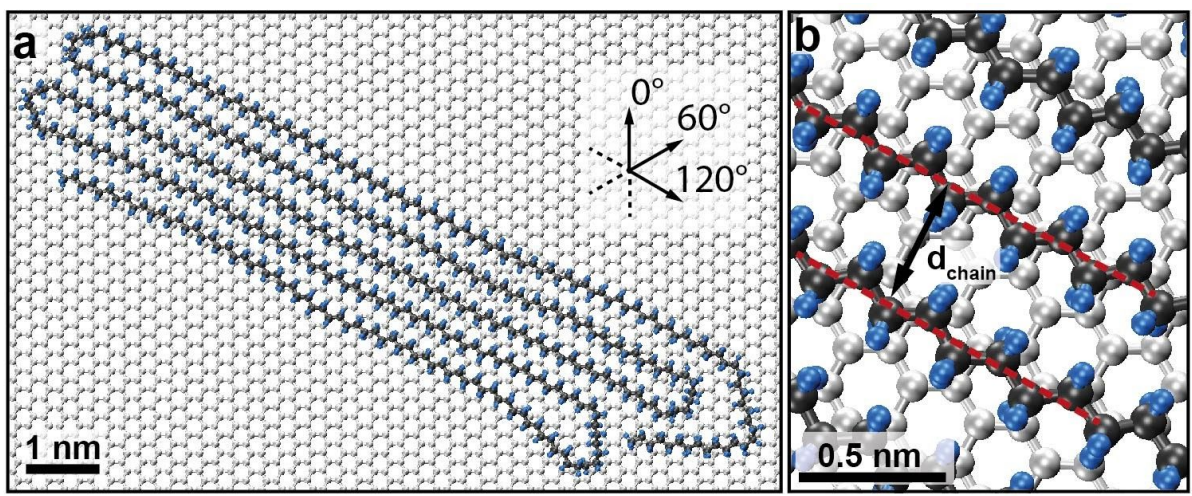

**Figure 1.** Equilibrium configuration of PE (C black, H blue) on graphene (light grey) at T = 290K. **a**: orientation of PE along principal directions of graphene (black arrows). **b**: Inter-chain distance of polymer corresponds to approx. double the graphene surface periodicity.

We begin by studying the equilibrium structure of a monolayer polymer crystal. Figure 1a displays the equilibrated conformation of a single PE molecule comprising 200 monomers, at a temperature T = 290 K. The formation of such a fully-crystalline structure, starting from a random amorphous polymer chains state, is a complex process that is taking place on a 100 ns time scale (see Supporting Information). Within the range of system temperatures employed during the equilibration, the polymer chain typically exhibits several linear segments with an average length of 35.4 (3.0) monomers, corresponding to a spatial extension of around 10 nm (for a detailed discussion of the linear segment length dependence, see Supporting Information). Chain folding ultimately results in the formation of a two-dimensional crystalline structure with an inter-chain distance of 4.61(6) Å at room temperature, close to double the periodicity of the underlying graphene (Fig. 1b).[26] Notably, the crystallite aligns itself along the principal directions (indicated by the black arrows in Fig. 1a) of the honeycomb-shaped substrate. The angle in between principal directions in 60°.

The orientational linkage of the adsorbed polymer crystallite on the graphene follows from the preferred location of its hydrogen atoms relative to the carbon hexagons. Specifically, the polymer's hydrogen atoms tend to be positioned in the hexagon center for energy minimization.[31] This is not continually possible, since the two crystalline structures are not fully lattice-matched (bond lengths and angles: 0.142 nm and 120 ° in graphene, 0.153 nm and 114° in PE, respectively), resulting in slight deviations from the preferred alignment along the length of the polymer folds.

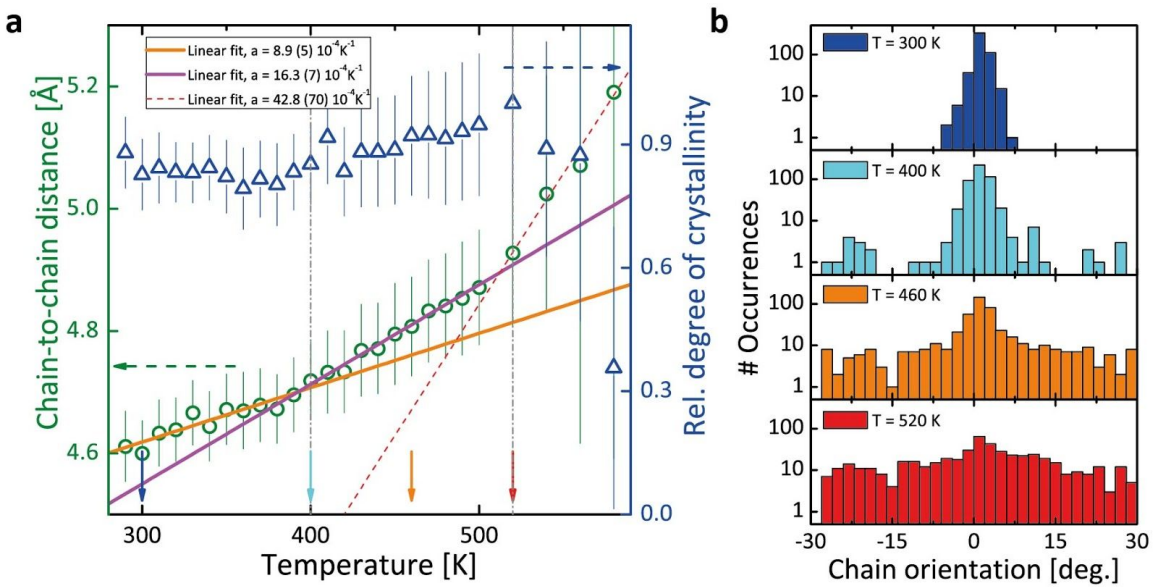

**Figure 2.** Polymer configuration in thermal equilibrium. **a**: Inter-chain distance (green circles, left y axis) and relative degree of crystallinity (blue triangles, right y axis) as function of polymer temperature. Increased thermal expansion above $T_1$ = 400 K and again above $T_2$ = 520 K (dashed-dotted grey lines), while relative crystallinity is largely maintained. Colored arrows indicate PE histogram temperatures in b. **b**: Distribution of crystallite orientations with respect to substrate`s principal directions (Fig. 1a, mapped to -30 deg to 30 deg), as given for T = 300 K, T = $T_1$ = 400 K, T = 460 K, and T = $T_2$ = 520 K.

Since vacuum is a bad solvent, the polymer adopts a compact, adsorbed state that resembles a 2D-pancake. The driving force of the polymer folding can be understood as a competition between gain in non-bonded potential energy for those PE atoms, which are coming closer by minimizing the boundary length of the 2D compact structure, the local increase in the dihedral energy due to a folding event, and the entropy of 2D, Hamiltonian-walk-like chain arrangements. Note that such a competition is taking place within a 2D hexagonal superstructure, that is imposed by the underlying substrate.

The inter-chain distance of the adsorbed polymer is strongly temperature-dependent[1]. We investigate this behavior in detail by sampling the inter-chain distance as well as the relative orientation of the 2D crystallite with respect to the substrate, in a series of MD simulations at different temperatures, over an extended period of time (N = 500 samples per temperature, in total 50 ns). Figure 2a shows the averaged chain-to-chain distance (green circles) as a function of the polymer temperature, which increases from about 4.61 Å at T = 290 K to 4.93 Å at T = 520 K.

Interestingly, the respective thermal expansion coefficient $a_T$ exhibits three different regimes. First, below a threshold temperature of about $T_1$ = 400 K, we find $a_{T1}$ = 9x10$^{-4}$K$^{-1}$. This value is considerably smaller than the bulk value for crystalline PE under comparable conditions, $a_{bulk}$ = 19(2)x10$^{-4}$K$^{-1}$.[32] Such a difference is not surprising: The strong interaction with the substrate serves as a structural template, impeding polymer movement and expansion[26] (energetic analysis found in the Supporting Information). Moreover, the single polymer chain can have a substantially higher crystallinity than its bulk counterpart, which further decreases the thermal expansion of the crystallite.[33-34]

In the second regime, between $T_1$ and $T_2$ = 520 K, the expansion coefficient nearly doubles to $a_{T2}$ = 16x10$^{-4}$K$^{-1}$, while the relative degree of polymer crystallinity remains constant (blue triangles, definition of relative crystallinity found in the Supporting Information). Finally, in the third regime, above the onset of superstructure melting, at T > $T_2$, $a_T$ rises steeply with decreasing internal order, until the inter-chain distance is no longer well-defined. Notably, the observed melting temperature of PE is strongly elevated compared to its bulk value ($T_{bulk}$ = 410 K).[35] This is a common feature for surface-induced crystallization of specific monolayer polymer systems as a result of strong polymer - substrate interaction.[36-38] Additionally, the monomeric structure of PE and the topography of the graphene are close to commensurate, guaranteeing very evenly-spaced folded-chain segments (see Fig. 1b).

The correlation between the inter-chain distance and the orientational distribution of the crystallite with respect to the substrate is examined in Fig. 2b. At four specific temperatures (indicated by the colored arrows in Fig. 2a), Fig. 2b displays the orientational distribution of the crystallite with respect to the substrate. Below $T_1$, we observe nearly perfect registration to the substrate (upper panel), which is increasingly lost for higher temperatures (center panels), until a nearly uniform orientational distribution is observed above $T_2$ (bottom panel). This finding strongly indicates that the increased thermal expansion beyond $T_1$ is facilitated by a deregistration of the polymer crystallite from the substrate. However, this deregistration is not immediately accompanied by an internal loss of order (for $T < T_2$), but instead results in a freely moving crystallite on the substrate with high in-plane mobility, thus representing a novel type of floating solid phase.[39-40] A more detailed analysis of this rotation is given in the Supporting Information.

---

[1] During the simulation, the graphene conformation ($T_{GR}$ = 293 K) has been kept fixed.

**Ultrafast Dynamics of 2D Crystallites**

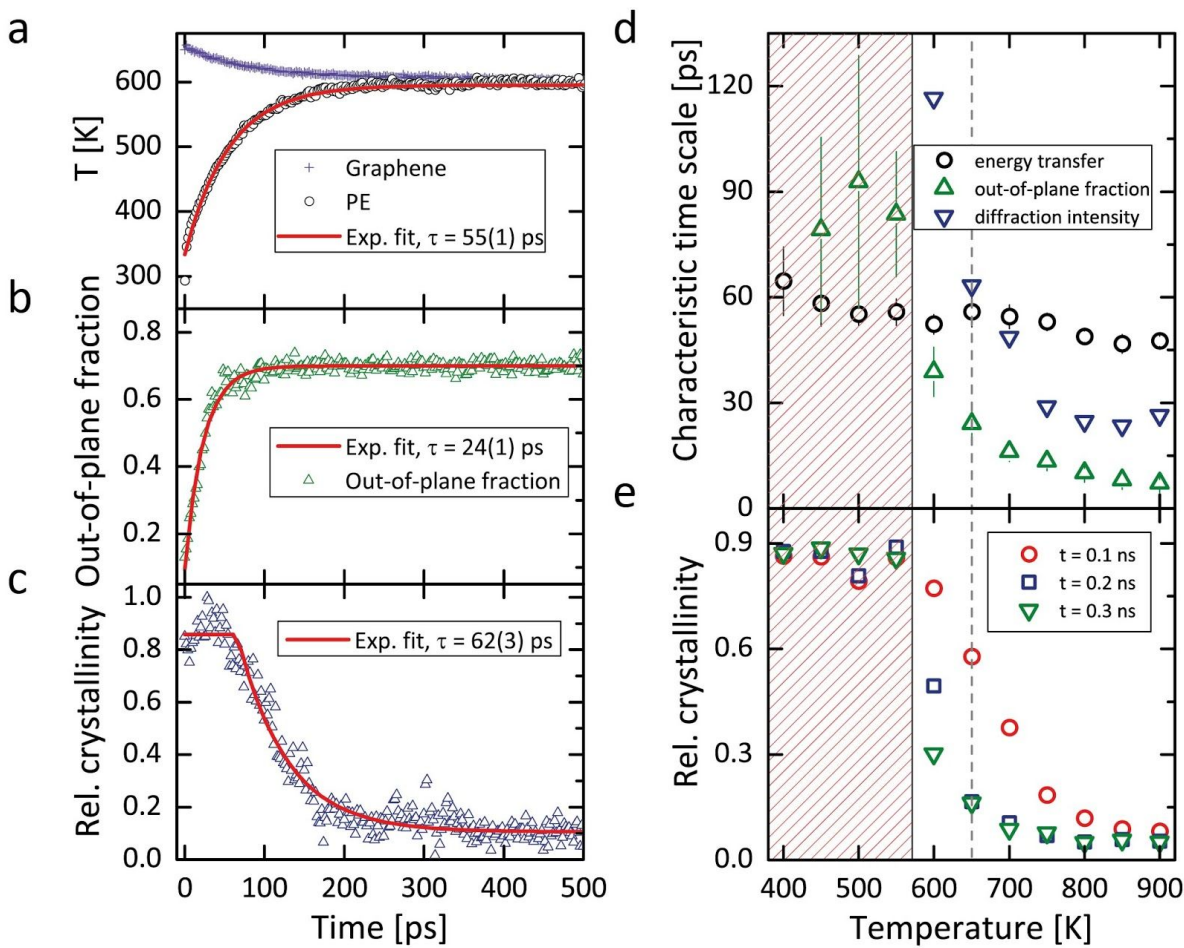

**Figure 3.** Temporal evolution of crystalline PE on heated graphene. Initial PE temperature $T_{PE}$ = 293 K. **a-c**: Initial graphene temperature $T_{Gr}$ = 650 K. **a**: Temperature of graphene (violet crosses) and PE (black circles). **b**: Fraction of PE out-of-plane dihedrals. **c**: Relative PE crystallinity. **d**: Time scales as found in a-c as function of $T_{Gr}$. **e**: Relative PE crystallinity as function of $T_{Gr}$ taken at t = 0.1 ns (red circles), t = 0.2 ns (blue squares), and t = 0.3 ns (green triangles). Dashed grey line indicates $T_{Gr}$ in a-c. Patterned red area: no substantial change in PE relative crystallinity.

Beyond the static properties of this complex heterosystem, we are interested in the momentum and energy transfer between graphene and polymer, which govern, for example, molecular friction and thermal boundary resistance. These are accessible by studying the structural response of the ordered polymer on the ultrafast time scale, after rapid thermal excitation of the substrate. We thus perform non-equilibrium NVE simulations on flexible graphene. Specifically, we analyze the crystallite`s behavior upon fast energy transfer from the graphene by assuming a strongly increased initial temperature of the substrate only, as it may be caused by, *e.g.*, pulsed laser excitation. In Figs. 3a-c, we plot a number of dynamical quantities as a function of time at a fixed initial graphene temperature of $T_{Gr} = 650$ K. Their respective temperature dependences are displayed in Figs. 3d-e. Specifically, Fig. 3a shows the (kinetic) temperature distribution for graphene (violet crosses) and PE (black circles). We find the energy transfer time from substrate to polymer to be on the order of 50 - 60 ps, only weakly depending on the initial graphene temperature (Fig. 3d).

The temperature equilibration is accompanied by a rapid increase (in about 24 ps) of the fraction of polymer dihedrals, which are out-of-plane with respect to the substrate (Fig. 3b). In order to evaluate the local order of the folded-chain structure, we compute the intensity of the polymer superstructure peak in a 2D Fourier transformation (Fig. 3c). This leads to a substantially larger characteristic time scale of about 60 ps. Notably, the loss of order only sets in after a certain threshold temperature of $T \cong T_2$ is exceeded (see also Fig. 2a). In stark contrast to the energy transfer time, these two features are very sensitive to changes in the initial substrate temperature. Specifically, their characteristic time scales are decreasing by a factor of about six upon a temperature increase of 50 % (Fig. 3d).

Structurally, we observe a strong change in the dihedral concentration for trans and gauche conformations on time scales comparable with that of polymer heating: The fraction of trans dihedrals decreases from initially 95 % to just below 70 % while the gauche concentration rises from about 3 % to 17 % in about the same time (Fig. 4a). Notably, both curves do not exhibit a plateau region such as the relative crystallinity in Fig. 3c. Instead, their time scales are directly linked to the temperature of the polymer.

Figure 4b displays the radial distribution function (RDF) of the PE backbone for different times between 0 ps and 500 ps without contributions from the four nearest neighbors along the chain. The initial RDF displays three peaks, which are related to the distances as shown in the inset. Notably, the highest peak, (iii), originates from a superposition of several similar interatomic distances. The center peak, (ii), corresponds to the inter-chain distance as previously described (cf. Fig. 2a).

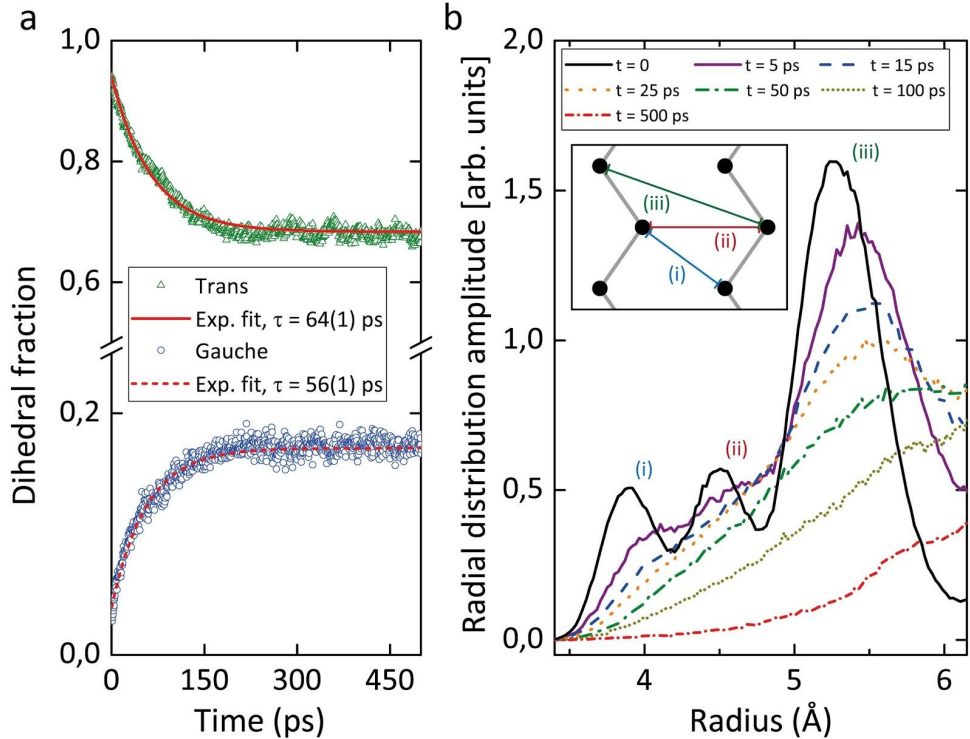

**Figure 4.** Structural evolution of PE at T = 650 K. **a**: Fraction of dihedrals in trans (green triangles) and gauche (blue circles) conformation, respectively. **b**: Radial distribution function of PE for different times, averaged over 10 ps (± 5 ps). Peak labels (i, ii, iii) refer to dominant distance contributions as depicted in inset.

Upon heating (initially, $T_{Gr}$ = 650 K), the peak's amplitude substantially decreases and shifts to larger radii on a few-picosecond time scale. Notably, peaks (i) and (ii) reduce their amplitude on a somewhat shorter time scale compared to (iii). This could be explained by the very rapid increase of dihedrals out-of-plane with respect to the graphene (Fig. 3b): upon backbone rotation, distances (i) and (ii) experience a stronger relative distortion than (iii), resulting in a fast decrease of the corresponding peaks. On the other hand, this shifts intensity to the broader (iii) peak, which is less sensitive to this rotation, leading to a slowed decline in its amplitude. Possibly, also the slight incommensurability between the solid support and the ideal 2D floating-solid structure could impart stress and predominantly distort the distances (i) and (ii).
8

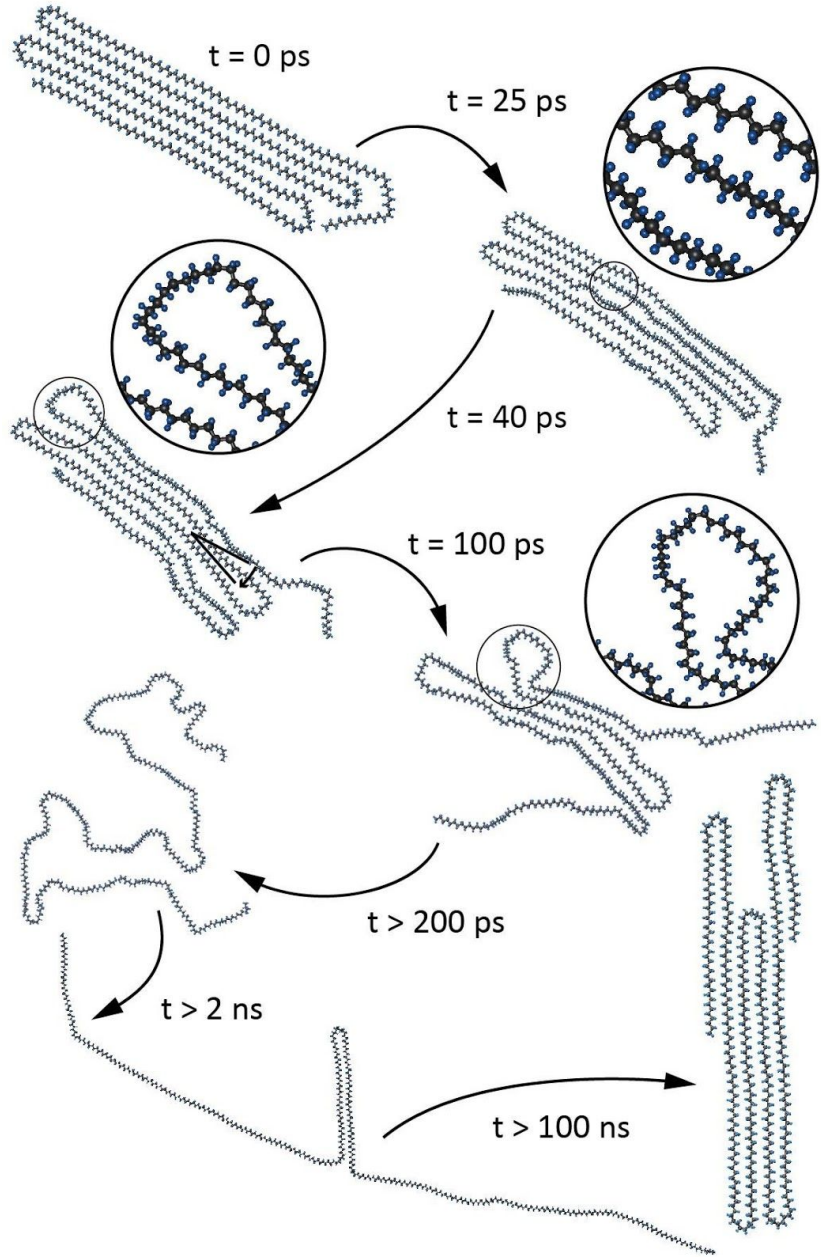

**Figure 5.** Atomistic sketch of PE superstructure melting. Initial temperatures $T_{PE}$ = 293 K, $T_{Gr}$ = 650 K. t = 0 ps: Initial PE folded-chain crystalline conformation with orientational linkage to substrate. t = 25 ps: Substrate-adsorbate energy transfer, segment breakdown, PE dihedrals rotate out-of-plane, floating solid phase. t = 40 ps: Loop formation at folds, crystallite rotation. t = 100 ps: Mobile regions move away from crystallites center, loss of order. t > 200 ps: Completely amorphous conformation. t > 2 ns: Re-registration with substrate. t > 100 ns: Full recrystallization, comparable to initial state.

**Physical Picture & Discussion**

Considering the above discussion, a general physical picture of the non-equilibrium single-chain crystal polymer dynamics can be derived as follows (Fig. 5): Initially, the polymer is equilibrated in a folded-chain crystalline superstructure with orientational linkage to the graphene. Upon rapid heating of the graphene substrate, energy is transferred to the adsorbate on a 50 ps - 60 ps time scale. In parallel, a large fraction of the polymer backbone rotates out-of-plane, resulting in a deregistration of the crystallite from the substrate. However, only a small change in orientation is visible on the short time scale (t = 25 ps). Notably, the formation of the floating phase is driven by internal changes in the polymer conformation, which lead to a change in adhesive interaction (free energy) with the substrate, and can therefore not be directly described by the classical one dimensional Frenkel-Kontorova model.[41]

The increased mobility of the molecular chain then allows for the creation of increasingly large loops at the folds (t = 40 ps). When these now mobile outer regions move away from the crystallites center, we observe a loss of crystallinity (t = 100 ps). After about 200 ps, this results in an amorphous state with no discernable correlation to the substrate order. Up to this point, the process can be interpreted as a heterogeneous (from the outside in) superstructure melting *via* an intermediate floating solid phase.

The segmental relaxation time of a polymer is defined through the time-correlation function of a vector at the monomer level, such as the dipole moment. Notably, all above characteristic time scales are much smaller than the segmental relaxation time of polymer chains of a free-standing thin PE film supported by graphene at equilibrium, in the vicinity of the polymer/graphene interface.[8,12,13] For the latter, strong dynamical heterogeneities due to polymer/graphene and polymer/vacuum interfaces have been observed. Our findings clearly demonstrate the impact of the molecular conformation on the dynamical behavior in two-dimensional single-chain crystal structures compared to interlaced, multi-chain polymer thin films supported by the same substrate (here graphene), due to both the additional free volume in the single-chain systems and the strong temperature dependence of the chain mobility. These result in (a) strong anisotropic chain mobilities (different friction for the in-plane and out-of-plane dynamics) at the single-chain polymer/graphene interface, and (b) faster dynamics (lower friction) compared to typical many-chain polymer thin films.

Interestingly, the described loss of order of the adsorbate is reversible: Upon a decrease of the system temperature, the molecule is nearly instantaneously re-registered to the substrate (t > 2 ns), which ultimately facilitates a full recrystallization on a 100 ns time scale (computation see Supporting Information). The final conformation again shows registration as well as a high degree of internal order comparable to the initial one, although not with the same orientation or folding.

Whereas nanostructured polymeric systems such as the one described in this work are inherently challenging to investigate experimentally, very recent experimental evidence supports the above findings.[20] In particular, ultrafast low-energy electron diffraction (ULEED) was employed to record the superstructure disordering dynamics of poly-(methyl methacrylate) (PMMA) adsorbed on free-standing graphene with picosecond temporal resolution. Despite the higher degree of structural complexity of PMMA and its larger molecular weight compared to PE, we find qualitative agreement on the structural level and regarding the time scales observed, namely (1) the formation of monomolecular polymeric layers of folded-chain crystallites with superstructure periodicity close to double that of the substrate, (2) an orientational linkage to the substrate, (3) comparable time scales for energy transfer and superstructure melting, (4) reversibility of the process regarding correlation lengths and long-range order. Experimentally, however, fluence-dependent time constants were not yet resolved.

The current work is able to track the actual pathway of melting of a crystalline polymer monolayer adsorbed on graphene: The data presented here strongly indicate a two-step superstructure melting process via the formation of a floating solid phase. Additionally, it provides an atomistic view on the trajectory of a single polymer chain during the transition, as well as the relevant time scales for energy transfer across the bilayer. Our findings are directly related to the non-equilibrium (anisotropic) molecular friction at the polymer/graphene interface. The time-resolved study highlights the detailed molecular motions involved in the formation of the floating solid phase and its correlation with the overall registration and orientation of the crystalline polymer with the substrate. In particular, we find the characteristic time scales for the ultrafast melting of the single-chain crystal to be much smaller than the equilibrium segmental relaxation times of multi-chain polymer thin films.

Our study yields a fundamental, atomic level understanding of the structure and dynamics of a single polymer chain in contact with graphene, under equilibrium as well as non-equilibrium conditions. We find a striking manifestation of a phase unique to two-dimensional systems, namely a floating solid, which is -- perhaps surprisingly -- also appearing as a transient structure on the few picosecond time scale. At this point, it remains open, whether the occurrence of this phase is a universal phenomenon or if it is a strong function of the polymer or interface properties. More generally, we believe that the rich behavior of this model system will stimulate further studies on the dynamics of multi-phase hybrid systems under non-equilibrium conditions.

## Methods

We performed all-atom (AA) molecular dynamics simulations in the GROMACS simulation environment. All subsequent analysis of the simulation data as well as the initial molecular design was performed with MATLAB. Within the simulation, we employ the OPLS-AA force field model[42-43] with periodic boundary conditions to model the polymer. The detailed force field parameters are given in Tab. 1 in the Supplementary Materials.

The graphene is initially modelled as a set of Lennard-Jones carbon atoms, centered and subsequently fixed at their crystallographic positions (lattice constant is 2.462 Å at T = 300 K) in the xy-plane. The graphene flake size is chosen large enough to not affect the polymer conformation (between 7 - 60 nm edge length). For the initial equilibration, we performed replica exchange (RE) MD simulations for a temperature range between 250 and 400 K (about 10 % exchange probability within 2 ps, 100 ns run time). For the orientational analysis, the equilibrated PE conformations are taken and their trajectories recorded at temperatures between T = 290 K and 600 K (50 ns runtime, no RE employed). The folded-chain orientation and relative crystallinity are determined by Fourier transformation of the polymer backbone conformation projected on the xy-plane.

In case of the non-equilibrium simulations of the system, we perform NVE computations of the equilibrium polymer conformation adsorbed to flexible graphene. For the initial conformation, the temperature of the latter is stepwise increased and equilibrated for 0.5 ps, while the polymer temperature remains fixed.

AUTHOR INFORMATION

**Present Addresses**
†Fraunhofer Institute for High-Speed Dynamics, Ernst-Mach-Institut, 79104 Freiburg, Germany

**Author Contributions**
The manuscript was written through contributions of all authors. All authors have given approval to the final version of the manuscript.

ACKNOWLEDGEMENT
We gratefully acknowledge support by the Deutsche Forschungsgemeinschaft (DFG) (SFB1073 "Atomic Scale Control of Energy Conversion", Projects A03 and A05). This work was funded by the European Union (EU) within the Horizon 2020 ERC-StG. "ULEED" (ID: 639119).

ABBREVIATIONS

MD, molecular dynamics; PE, polyethylene; RDF, radial distribution function; ULEED, ultrafast low-energy electron diffraction; PMMA, poly-(methyl methacrylate); OPLS, optimized potentials for liquid simulations; AA, all atom.

**Supporting Information Available:** rotation of the polymer crystallite, initial adsorption and crystallization, energy evolution during adsorption and crystallization, time scale of recrystallization, definition of relative crystallinity, comparison of united-atom and all-atom simulations, linear segment length dependence, influence of the substrate size, OPLS-AA force field parameters.

# Supporting Information

# Dynamics and Structure of Monolayer Polymer Crystallites on Graphene

Max Gulde,*,† Anastassia N. Rissanou,‡ Vagelis Harmandaris,*,‡,§ Marcus Müller," Sascha Schäfer,† Claus Ropers†

†4th Physical Institute - Solids and Nanostructures, University of Göttingen, 37077, Göttingen, Germany
‡Institute of Applied and Computational Mathematics, Foundation for Research and Technology Hellas, 71110 Heraklion, Crete, Greece
§Department of Mathematics and Applied Mathematics, University of Crete, 71409, Heraklion, Crete, Greece
"Institute for Theoretical Physics, University of Göttingen, 37077, Göttingen, Germany

**Corresponding Authors**
*E-mail: max.gulde@gmail.com
*E-mail: harman@uoc.gr

**Present address**
(M.G.) Fraunhofer Institute for High-Speed Dynamics, Ernst-Mach-Institut, 79104 Freiburg, Germany

**Rotation of the polymer crystallite**

Figure S1 illustrates an atomistic picture of a molecular rotation. Specifically, Fig. S1A displays the average distance between the PE backbone and the graphene[2] (blue line) as well as the absolute rotational angle (orange line) as a function of time. In this particular case, the crystallite performs a single counter-clockwise turn of 60°. Colored arrows indicate the points in time of snapshot conformations as depicted in Fig. S1C. The total fraction of polymer dihedrals rotated out-of-plane (blue line) with respect to the graphene substrate and the relative degree of crystallinity (orange line) are given in Fig. S1B.

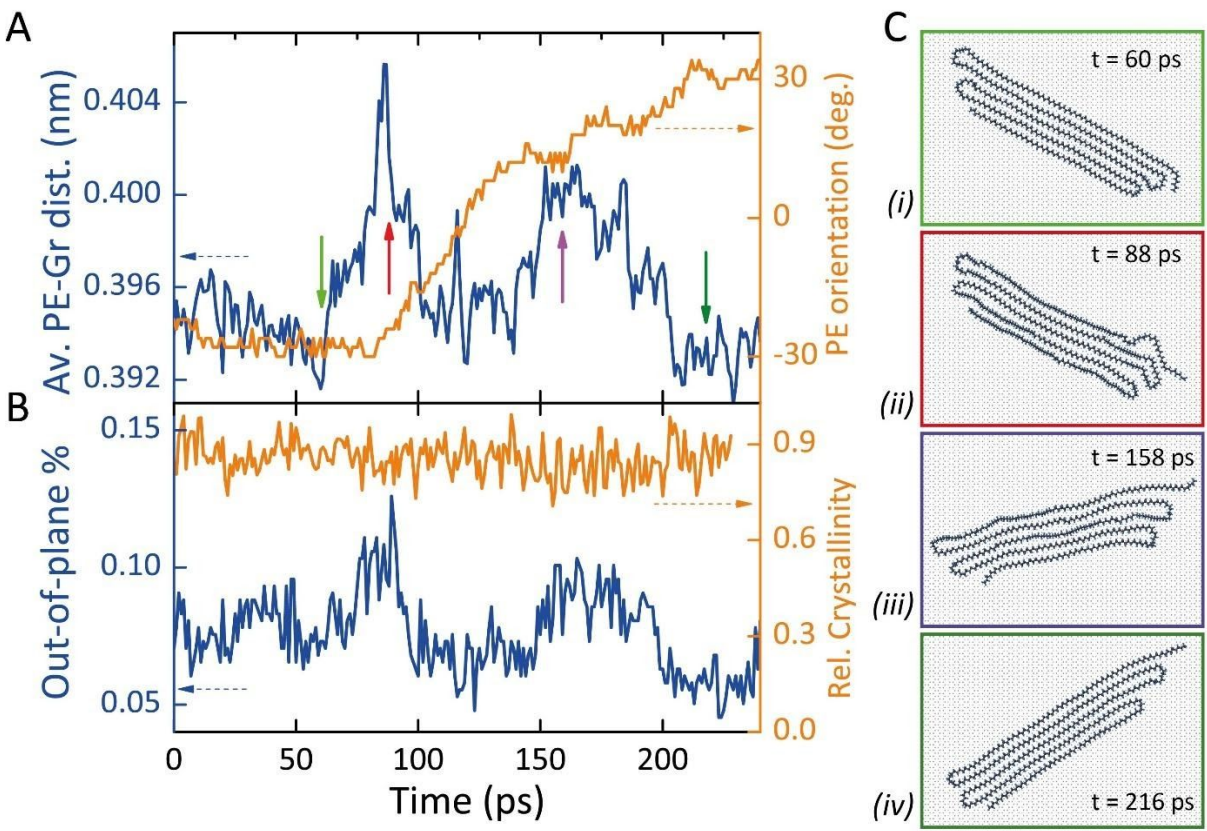

**Figure S1.** Rotation of PE crystallite at T = 500 K. **A**: Mean PE backbone to graphene distance (blue line) upon rotation of the crystallite (orange line). Colored arrows indicate points in time for conformational snapshots in C. **B**: Relative frequency of out-of-plane PE dihedrals (blue line) and relative degree of crystallinity (orange line). **C**: Snapshots of PE conformation on graphene at different points in time. PE molecule moves as a unit. For rotation, PE dihedrals rotate out-of-plane with respect to graphene.

---

[2] Graphene position has been held fixed during this simulation.

The rotation takes place roughly at times 80 ps < t < 210 ps. The rotational speed is not constant over the whole rotation. Instead, between 80 ps and 150 ps, the molecule rotates by 40-45°, whereas for last 15-20° the speed is approximately halved. In between, the molecular rotation is stopped for few tens of picoseconds.

Notably, both partial rotations commence with an increase in PE-graphene distance by about 0.1 Å at t = 80 ps and 0.06 Å at t = 150 ps. The spatial separation is accompanied by an increase in the fraction of PE dihedrals, which are rotated out-of-plane with respect to the graphene. However, this features is not constant over the duration of the orientational change, but lasts only 20 – 50 ps. In total, the rotation has a characteristic time scale of about 150 ps. Interestingly, during this time, the relative degree of crystallinity does not decrease visibly (Fig. S1B). In particular, the crystallite rotates as a unit (Fig. S1C), opposite to turning by formation of kinks along the segments, separating the crystallite into several parts with different orientation. This can again be understood as the result of an energetically favorable self-proximity, holding the crystallite together during a rotation, even when close to the melting onset of the system.
Additionally, this observation further corroborates the idea of a floating solid phase. As stated in the main text, this was not observed experimentally on a similar system, comprising a PMMA superstructure adsorbed to free-standing graphene.[20] It is therefore to assume that the characteristic time scale of a rotation depends strongly on the complexity and the length of the polymer as well as the overall degree of substrate coverage. Hence, the observed time scale in the observations should be best understood as a lower limit for the rotation time of a polymer crystallite.

**Initial adsorption and crystallization**

Initially, the polymer is constructed with a randomized backbone conformation and placed within the immediate vicinity of the graphene substrate (Fig. S2A). Partial adsorption of the nearby polymer chain occurs on a few-picosecond time scale. At around 1 ns, a significant part of the polymer chain is adsorbed and starts folding itself. However, this process is not limited to the polymer on top of the graphene, but also occurs on the freely floating end (Fig. S2B). Since the latter does not suffer from surface constraints, the folding process occurs more quickly. After about 10 ns, the adsorption, the folding into a crystalline structure as well as the orientational linkage of the polymer with respect to the graphene is nearly complete (Fig. S2C, cf. Fig. S3). However, smaller conformational "jumps" can still occur on later times, as can be seen in Fig. S2D, which displays the relative crystallinity as a function of time.

Interestingly, the described initial crystallization process is a lot faster than the recrystallization process out of the amorphous phase. This can be explained qualitatively by the higher rigidity of the adsorbed molecule during cool-down compared to the freely-moving initial conformation.

Specifically, whereas the partial adsorption depicted in Fig. S2 leads to a fast folding of the molecule, the amorphous molecule already close to the graphene is adsorbed in the momentary conformation, once the temperature is reduced. As the temperature then further decreases, rotation and folding of adsorbed chain segments becomes increasingly unlikely, ultimately leading to larger recrystallization times.

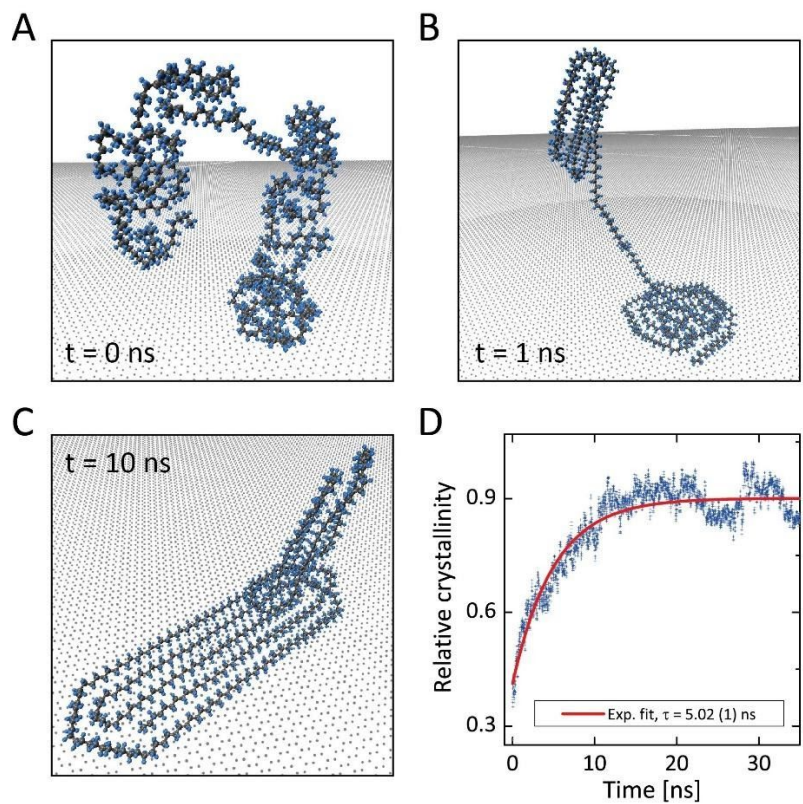

**Figure S2.** Initial crystallization process of single PE 200-mer. **A**: Random polymer configuration in substrate proximity. **B**: Adsorption to graphene, folded-chain configuration in adsorbed and not adsorbed polymer regions. **C**: Close-to-equilibrium folded-chain configuration, registration and orientational linkage to substrate. **D**: Relative degree of crystallinity as function of time averaged over T = 290 K to T = 350 K.

Notably, the linear segment length is independent on the initial PE chain structure: Firstly, recrystallized structures from the amorphous phase are qualitatively similar to the initial folded-chain configuration formed from a randomized, non-adsorbed polymer chain. Secondly, for a partly crystalline PE chain in vacuum, the energy gain during adsorption to the graphene substrate is much larger than the additional energy needed to sustain a slightly compressed folded-chain conformation (see section "Energy evolution during adsorption, crystallization, and in the floating solid phase"). Hence, the final configuration is not dictated by the kinetics of the adsorption process.

**Energy evolution during adsorption, crystallization, and in the floating solid phase**

The graphene substrate serves as a strong template for the PE conformation. In particular, the hydrogen atoms will try to minimize their conformational energy by arranging themselves in the center of the graphene hexagons as for example reported in.[31] Figure S3 displays the change in nonbonded energy as a function of time for three different temperatures. Whereas the energy gain by adsorption to graphene lies between 3.8 - 5.0 MJ/mol depending on the temperature (Fig. S3 A), the energy loss connected to conformational changes is substantially smaller around 1.5 - 2.0 MJ/mol (Fig. S3 B). Hence, adapting its interchain distance to the graphene template still results in a net energy gain for the polymer. For comparison, the inset in Fig. S3 B displays the energy gain from crystallization of the same polymer chain in vacuum.

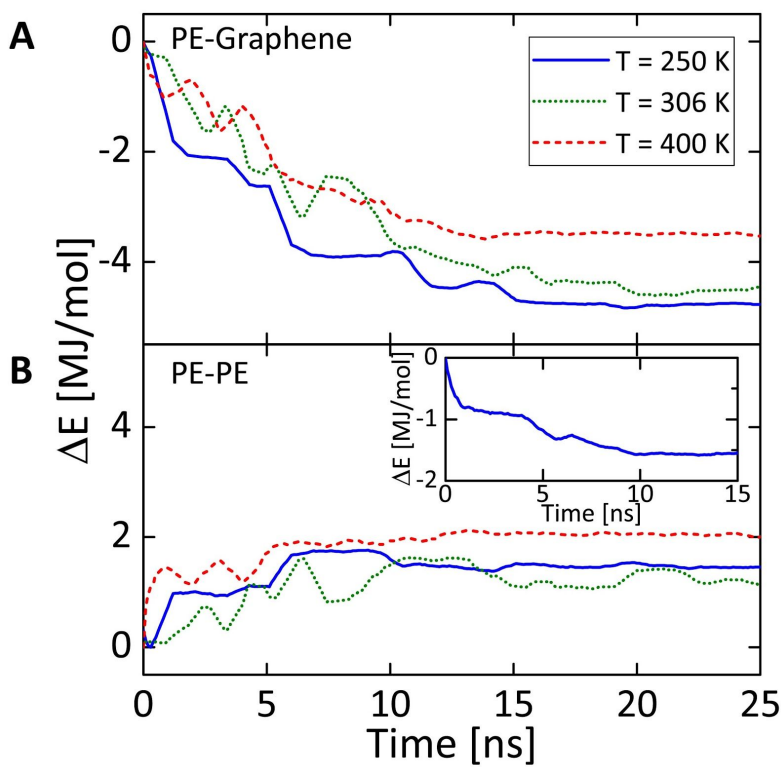

**Figure S3.** Nonbonded change in energy as a function of time and system temperature during crystallization and/or adsorption of PE (400-mer). **A**: Energy gain via adsorption of PE onto graphene. **B**: Additional conformational PE energy due to smaller lattice parameter. Inset: Conformational energy gain of PE in vacuum via crystallization.

Furthermore, we have investigated the temperature-dependent interaction energies within the polymer and between PE and graphene during the transition from the registered polymer crystallites to the floating solid phase and beyond. Figure S4 displays the interaction energy in both cases as a function of the system temperature in equilibrium. The change in energy in polymer-graphene and polymer-polymer interactions basically parallels the change in interchain separation as a function of temperature as depicted in Fig. 2 in the main text. Specifically, we observe a steeper rise of interaction energy with increasing temperatures: The specific change in interaction energy between PE and graphene increases by about 60 % upon transition to the floating solid phase (at around 400 K) from around 1 kJ/mol/K to approximately 1.7 kJ/mol/K. Similarly, the specific PE-PE interaction energy increases by around 75 % from 1.5 kJ/mol/K to around 2.6 kJ/mol/K. The steepened rise in interaction energy corroborates the loss of registration to the substrate above $T_1$. Notably, the ascending slope further increases at temperatures beyond $T_2$, when the crystallites become amorphous.

Comparable to the case of interchain separation, the transition from the solid to the floating solid phase and, at even higher temperatures, to the amorphous state, displays a pronounced signature in the interaction energy.

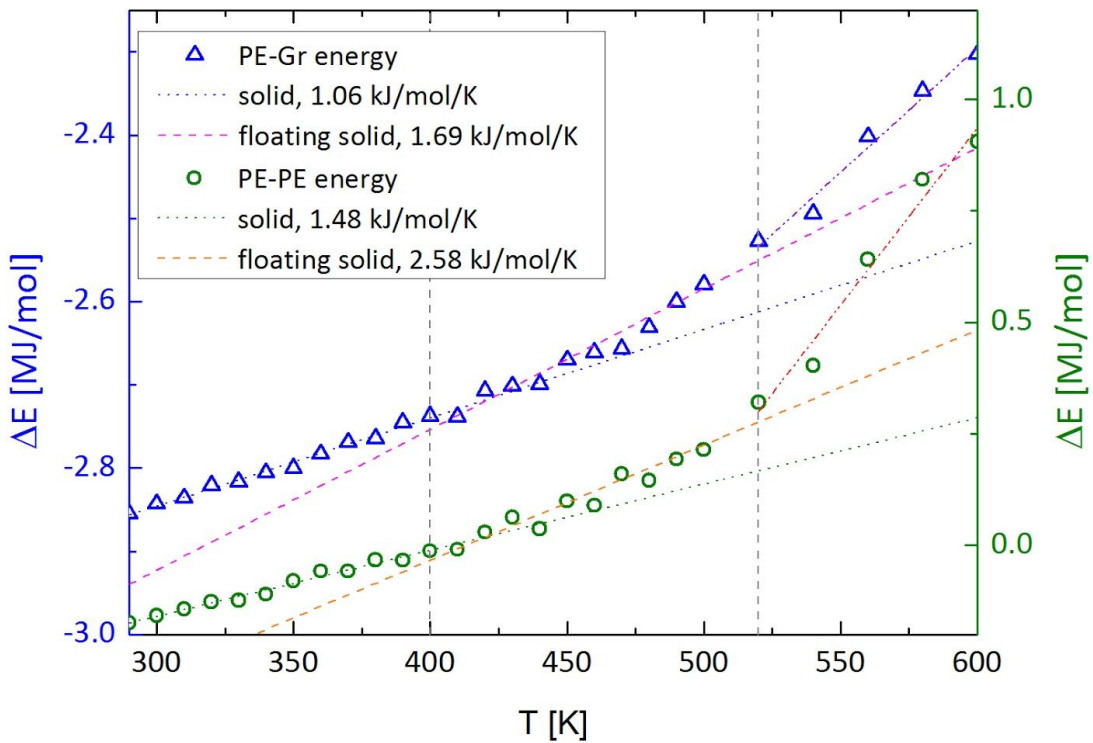

**Figure S4.** Change in polymer-polymer (green dots) and polymer-graphene (blue triangles) interaction energy as a function of temperature in equilibrium. The dashed and dotted lines represent linear fits for the temperature ranges 290 K < T < $T_1$, $T_1$ < T < $T_2$, and T > $T_2$ with the transition temperatures $T_{1,2}$ illustrated by gray dashed lines as in Fig. 2 in the main text.

**Time scale of recrystallization**

The recrystallization time of the polymer on top of the substrate is orders of magnitudes larger than the melting time. When one wants to access the respective time scale via molecular simulation, this potentially results in a tremendous computational effort, considering the number of atoms involved in combination with the amount of different initial configurations needed for acceptable statistics.

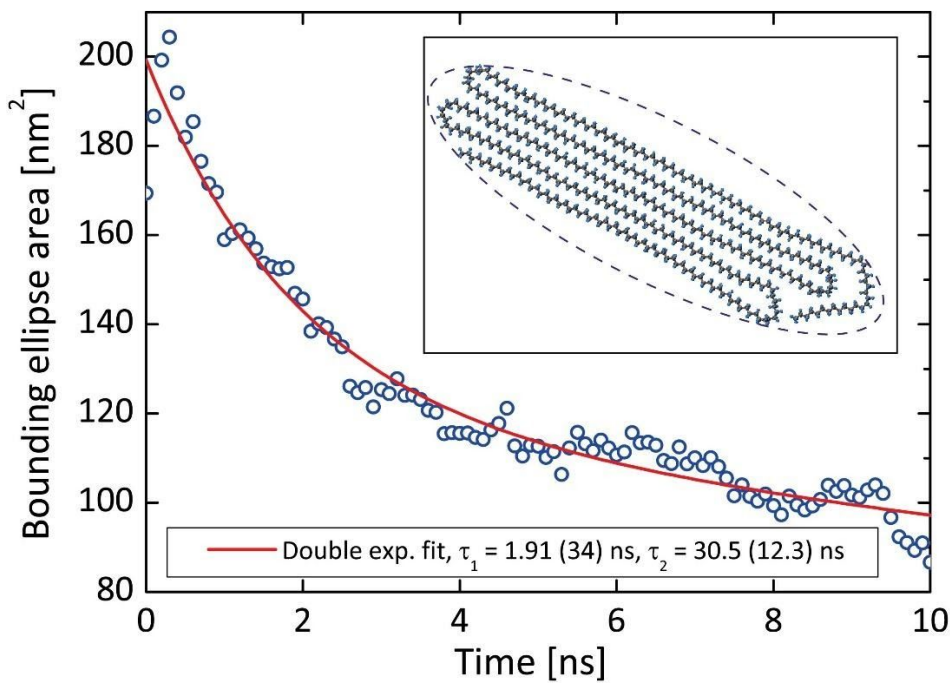

**Figure S5.** Crystallite compactness via bounding ellipse method as function of time averaged over 56 different initial states. Inset: exemplary bounding ellipse for compact equilibrium conformation.

Therefore, to nevertheless get an estimate for the recrystallization time of the superstructure from its amorphous phase, two simple models have been employed. First of all, eight different, randomly chosen initial configurations of amorphous PE on graphene have been extracted from the non-equilibrium runs at late times. Each of these has been equilibrated at seven different temperatures between 290 K and 350 K (equal distance). Subsequently, we computed the polymers conformational trajectory over a time span of 10 ns (hence total simulated time span of 560 ns).

Firstly, all resulting trajectories have been analyzed in terms of folding events and temporal evolution of the minimal bounding ellipse as a figure of merit for the conformations compactness. Specifically, more than 90 % of all recorded chain-foldings happen within the first 2 ns, before the polymer has completely registered to the substrate. From the number of folding events after t

= 2 ns, assuming a now roughly linear folding behavior, we derived an average time span of around 104 ns for the molecule to be entirely recrystallized. Complete recrystallization was defined as exhibiting the same number of folds as the average equilibrium conformation (4.7 folds).

In a second approach, as shown in Fig. S5, we fitted the conformational compactness of the adsorbed polymer[A1] with a double exponential decay function, resulting in a fast component ($t_1$ = 2 ns) and a slower component ($t_2$ = 30 ns). The constant offset was chosen to be the averaged area of the equilibrium conformations (Fig. S5, inset) of around 35 $nm^2$. Subsequently, we defined the recrystallization as complete, when the bounding ellipse area is within 10 % of the average equilibrium area, which is the case after 98 ns.

Naturally, the recrystallization will be strongly depending on many other parameters, including the substrate coverage, or the molecular weight and type of the polymer. Therefore, the value of around 100 ns extracted here should be seen mainly as a rough average lower limit.

**Definition of relative crystallinity**

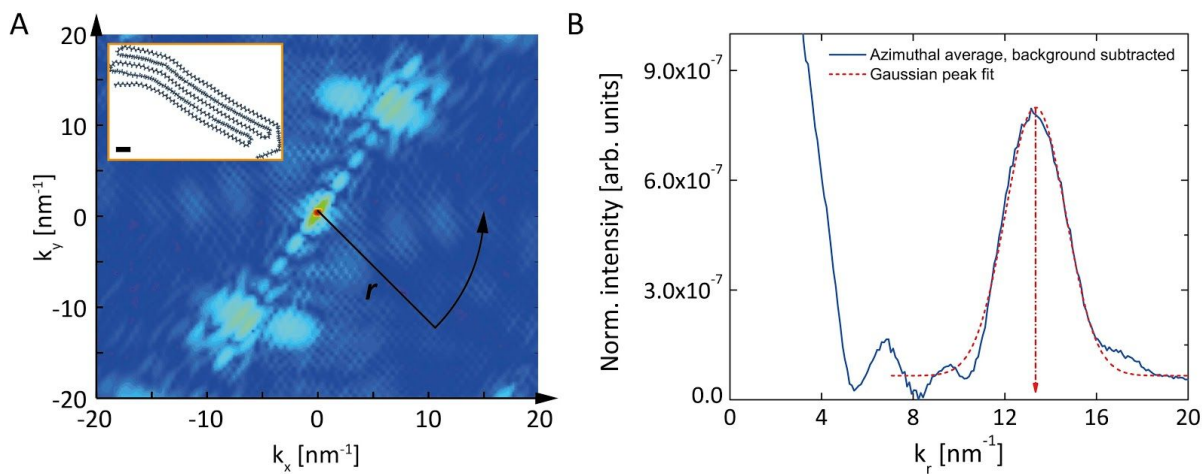

**Figure S6.** Determination of relative crystallinity. **A**: 2D Fourier transformation of polymer conformation (inset), subsequent azimuthal average along radius r. Black bar in inset has length of 1 nm. **B**: Gaussian fit of structure factor amplitude after background subtraction. Envelope height determines crystallinity, position of maximum the lattice constant.

The relative degree of crystallinity is defined via the structure factor amplitude of the polymer superstructure. First, we perform a 2D Fourier transformation of the polymer conformation only without substrate contributions (Fig. S6A and inset, respectively).

At this point, we can additionally determine the orientation of the crystallite by identifying the angle, under which the superstructure peak is strongest. To get the relative degree of crystallinity, however, the resulting pattern is azimuthally averaged starting from its center. This yields the scattering intensity as a function of the scattering angle $k_r$ (Fig. S6B). After subtracting an exponential background contribution, the peak is fitted by a Gaussian. The maximum of the Gaussian is taken as the figure of merit for the relative degree of crystallinity, while its position determines the lattice constant.

**Comparison of UA with AA simulations**

To investigate the orientational linkage of the polymer crystallites to the substrate, we performed united-atom (UA) simulations on a PE 200-mer as used in the all-atom (AA) simulations in this work. Figure S7A displays the orientational distribution of the polymer crystallite at comparable temperatures for both cases.

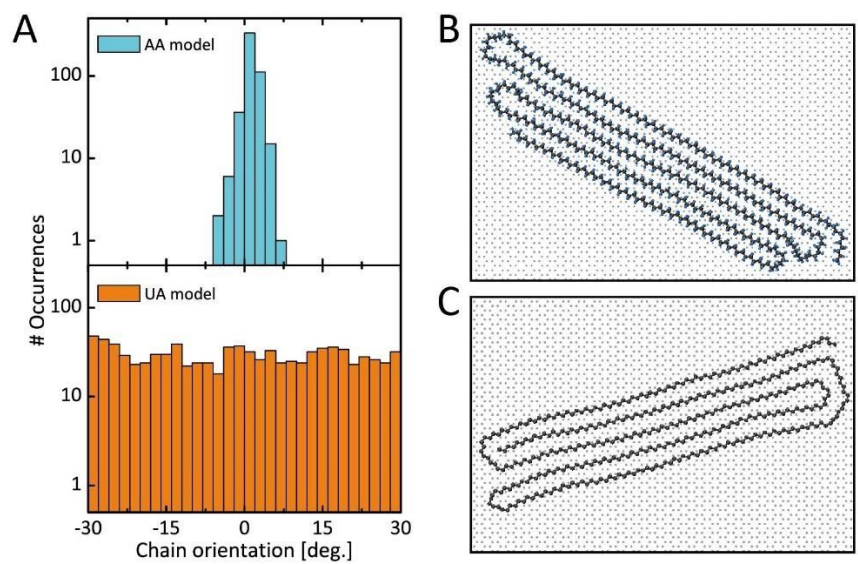

**Figure S7.** Comparison AA model and UA model. **A**: Distribution of crystallite orientations for AA model (top, T = 293 K) and UA model (bottom, T = 305 K). **B-C**: Exemplary crystalline conformation of AA (B) model with and UA (C) model without orientation linkage to the substrate.

For the UA simulations, we used the TraPPE-UA force field. It is apparent that the UA simulations show no linkage to the principle graphene directions. This follows from the lack of hydrogen atoms in comparison with the AA model. Specifically, the hydrogen atoms try to place themselves centrally within the graphene rings. This is most frequently possible, considering the bond length mismatch between the polymer backbone and the substrate, along the principal graphene directions.

**Linear-segment length dependence**

Firstly, the length of linear polymer chain segments has been analyzed with respect to the system temperature. Specifically, we performed REMD simulations at eight different temperatures between 250 K and 400 K and a total simulation time of 100 ns per temperature. Within this parameter space, only systems with temperatures below 310 K reached stable equilibrium conformation. A conformation was considered stable, if no significant change in its folded-chain conformation is observed. However, the molecule might rotate relative to the graphene substrate. At temperatures above 310 K, such a stable configuration was not reached within the observed time interval.

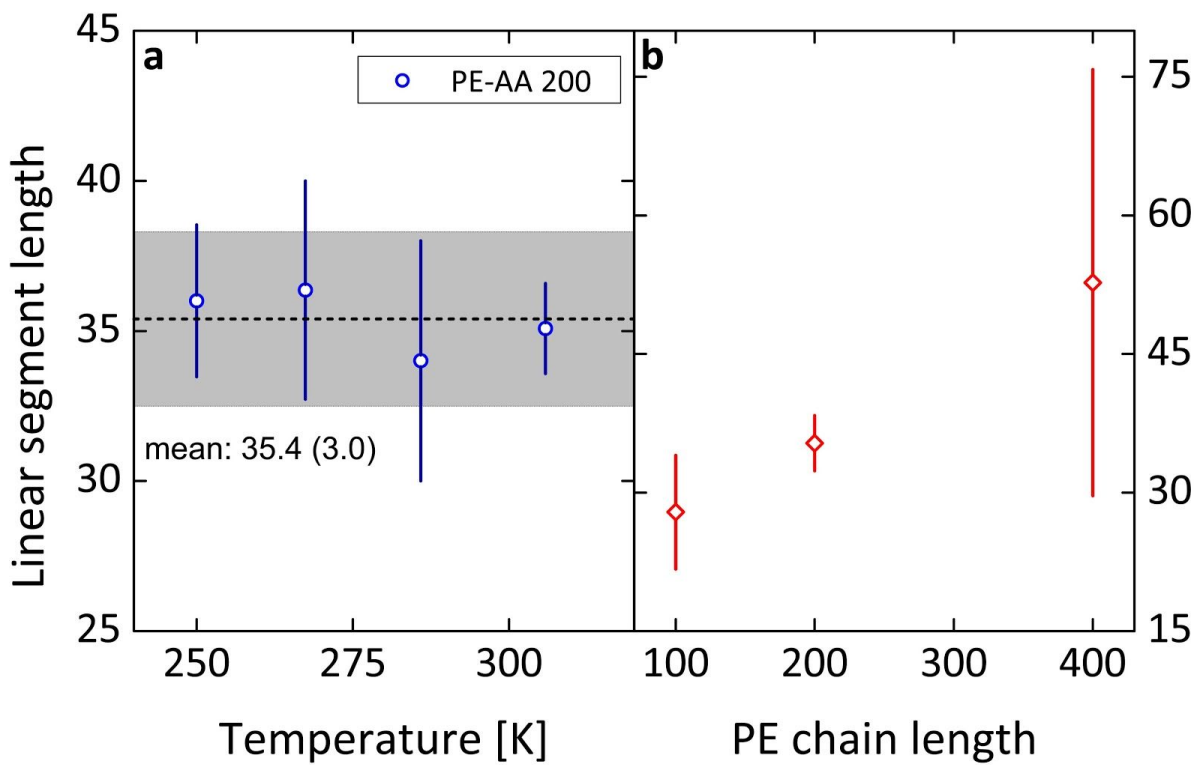

**Figure S8.** **A**: Average segment length as a function of system temperature for AA-PE with 200 monomers. N = 54 samples. Total average segment length and standard deviation indicated by black dashed line and gray area, respectively. **B**: Average segment length as a function of PE chain length in monomer units.

Figure S8 displays the average observed linear segment length of a PE 200-mer, as used for most of our simulations reported in this work. Within the analyzed temperature range and the recorded data, no significant dependence of the segmental length scale on the system temperature could be observed. The overall average segment length was found to be 35.4 (3.0) monomer units long.

It has to be noted, that unfinished segments, meaning segment which incorporated the beginning or end of the polymer chain, have not been included in this evaluation. Additionally, only segments parallel to the graphene orientation have been counted.

It has to be noted that the investigated temperature range is relatively narrow. We expect of course a reduction in linear segment length, once we pass the threshold below which we can achieve stable folded-chain crystal conformations.

Additionally, we investigated the average linear segment length as a function of total PE chain length (Fig. S8 B). Specifically, we investigated the monomer range between 100 and 400 monomers, finding an increase in average linear segment length with increasing monomer number. It has to be noted that the fluctuation in segment lengths is also substantially increasing with longer polymer chains.

**Influence of the substrate size**

The maximum length of a PE 200-mer chain fully extended in one direction is about 56 nm assuming 114° bond angles. We therefore chose a 60 nm sized, square sheet of graphene to perform the simulations of non-adsorbed PE with random configuration. Once the adsorption as well as the initial folding had occurred, simulations were continued on a square-shaped graphene flake with a edge length of 15 nm.

In order to avoid any influence of the substrate size, we verified for all simulations that a) the periodicity of graphene over the boundary was not violated and b) the polymer did not interact with itself over the periodic boundary. Additionally, we have not observed any conformational differences for larger simulation boxes. In fact, only for graphene flakes of below 10 nm and not completely crystallized polymer chains it has happened that the polymer was self-interacting over the boundary, resulting in conformations with either a single or no fold, very long segment lengths and no alignment to the substrate.

**Multichain polymer systems in 2D**

The investigation of systems comprising multiple polymer chains is of great importance. In 3D, a chain interacts with many neighbors (in the non-crystalline state) and there is a dramatic difference in chain configurations between a collapsed, isolated chain in a bad solvent, a Gaussian non-crystalline chain in a melt, and a chain in a 3D crystal. These differences are much less drastic in 2D because the chain in a bad solvent on a substrate adopts a rather compact, pancake-like configuration.

To demonstrate that the phenomena observed for a single chain are indeed universal and also valid in the case of interchain interaction, we performed simulations with multichain PE systems in the high packing limit[3]. Specifically, we analyzed a system comprising 11 collapsed PE chains (200 monomers each) on a 20 x 20 nm$^2$ graphene substrate (Fig. S9).

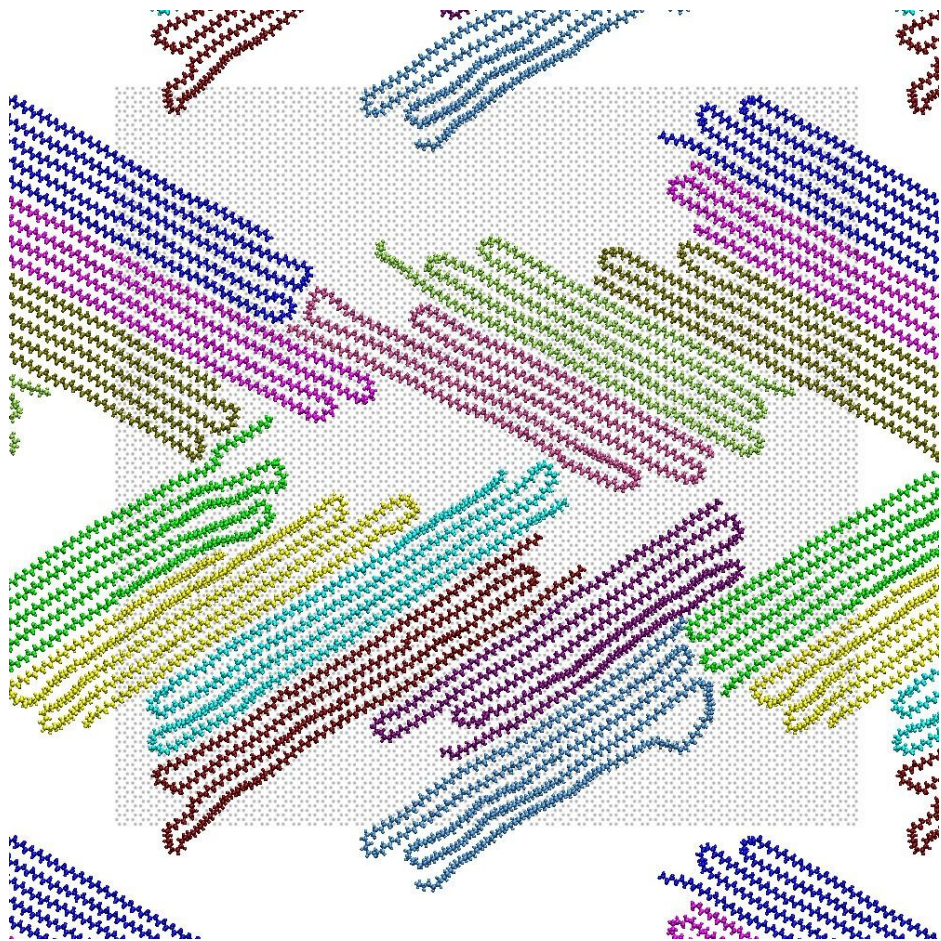

**Figure S9.** Equilibrated high packing PE system (colored) on graphene substrate (gray dots), periodic boundary conditions. T = 400 K.

As expected, the multichain system displays the same characteristics previously observed for the dilute case: (1) Individual polymers equilibrate in the described folded-chain crystal structure with 30 to 40 monomers per linear segment. Notably, interaction between neighboring chains exclusively occurs at the crystallite's perimeter without interweaving of single chain segments as expected for a 3D system. (2) The crystallites exhibit registration and hence orientational linkage to the substrate (compare Figs. 1 and 2). However, in a system comprising multiple chains, geometrical constraints could potentially result in the inability of individual crystallites to fully align with the substrate. (3) The manifestation of the floating solid phase is observed at temperatures between $T_1 = 420$ K and $T_2 = 500$ K, indicated by the enhanced increase in chain-to-chain distance (Fig. S10, regional linear fits displayed as solid lines). Notably, this feature is even more pronounced compared to the case of a single chain, as indicated by the dashed lines. Furthermore, the temperature range for the floating solid phase appears to be slightly smaller in the multichain system. However, these differences are within the statistical error. At $T_2$, the relative crystallinity rapidly drops as the polymer chains begin to melt.

---

[3] In the dilute case investigated in this work, the ratio between graphene atoms and PE atoms is around 7:1. In the high packing case, the ratio is close to 1:1.

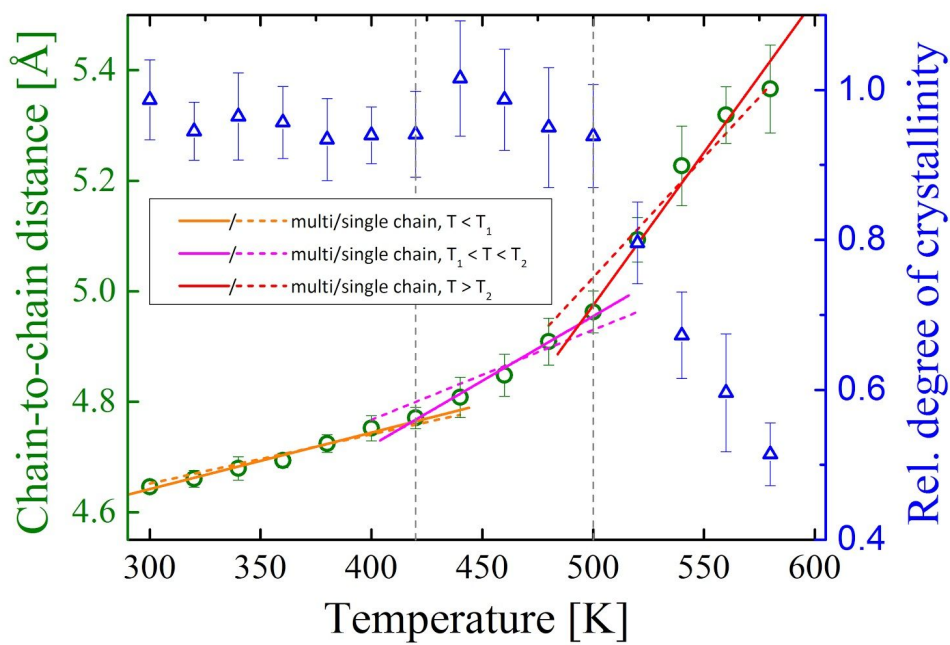

**Figure S10.** Temperature dependence of chain-to-chain distance (green dots) and relative degree of crystallinity (blue triangles) in multichain PE system. Floating solid phase region between gray dashed lines. Data fits in regions $T < T_1$, $T_1 < T < T_2$, and $T > T_2$ indicated by solid lines, respective slopes for single chain as dashed lines of same color.

These behavioral similarities between the dilute and the highly packed configurations result directly from the dimensionality of the systems: Even in the high packing limit, intramolecular interactions vastly dominate over packing interactions. The reason is found in vacuum being a bad solvent for the polymer: the polymer will try to keep individual chain segments close together. An energetically favorable conformation is hence achieved by minimizing the boundary length and adopting the aforementioned pancake-like configuration.

Generally, configurations in a 2D melt and isolated 2D "pancakes" scale in covered area proportional to N, the monomer number, while the length of the perimeter scales proportionally to sqrt(N). There are some differences in the shape of the boundary (fractal for full coverage,[A3] or smoother for slightly more isolated chains as in Fig. S9), but these do not affect the essential physical properties of the system, such as the local chain registration and deregistration, the thermodynamic properties or the nature and dynamics of phase transitions.

In particular, the short-time dynamical features described in our work, such as the dynamics of deregistration, occur on a more local level that is not affected by the overall shape of the polymer chain. In the limit of long chains, the interior of the pancake-like configuration dominates the behavior (boundary *vs.* area argument) and this holds true *a fortiori* for the short time dynamics where we do not expect a significant center-of-mass motion that could modify the interchain interactions.

**Comment on studies with PMMA**

In view of future studies investigating more complex systems, e.g., PMMA as in our previous experimental work, we believe that it is important to start with the simplest possible model system. Not only does the reduced computational effort in simulating a small number of atoms over relatively short time scales allow to quickly and completely sample each part of the system's intricate dynamics, we also believe that it will substantially facilitate the understanding of more complex systems in the future.

**OPLS-AA force field parameters**

The OPLS-AA force field was chosen, since it accurately describes the CH-pi interactions between the polymer and graphene.[A3] Additionally, it was found to be in good agreement with density functional theory.[26]

We used the following force field parameters throughout all performed GROMACS simulations (nonbonded function type = 1, combination rule = 3, pair generation, fudgeLJ = fudgeQQ = 0.5 as described in the GROMACS user manual 5.0.4).[40-41] Cutoff for all MD simulations was 1 nm.

Atoms

| parameter | H (PE) | C (PE) | C (Gr) |
|---|---|---|---|
| Mass [u] | 1.008 | 12.011 | 12.011 |
| Charge [eV] | 0.0 | 0.0 | 0.0 |
| Sigma [nm] | 0.25 | 0.35 | 0.34 |
| Epsilon [kJ/mol] | 0.1256 | 0.2763 | 0.2327 |

Bonds

| parameter | C-C (PE) | C-H(PE) | C-C(Gr) |
|---|---|---|---|
| Function type | 1 | 1 | 3 |
| $B_0$ [nm] | 0.1529 | 0.1090 | 0.1418 |
| $c_b$ [kJ/mol nm$^2$] | 224418.02 | 284709.43 | 478.90 |

Angles

| parameter | H-C-H | H-C-C | C-C-C (PE) | C-C-C (Gr) |
|---|---|---|---|---|
| Function type | 1 | 1 | 1 | 2 |
| $\theta_0$ | 107.8 | 110.7 | 112.7 | 120 |
| $C_{theta}$ [kJ/mol deg$^2$] | 276.33 | 314.01 | 488.61 | 562.2 |

Dihedrals

| parameter | H-C-C-C | C-C-C-C (PE) | C-C-C-C (Gr) |
|---|---|---|---|
| Function type | 5 | 5 | 3 |
| C1 [kJ/mol] | 0 | 5.443 | 25.12 |
| C2 [kJ/mol] | 0 | -0.209 | 0 |
| C3 [kJ/mol] | 1.256 | 0.837 | .25.12 |

**Table S1.** Force field parameters as used throughout the simulations.

ADDITIONAL REFERENCES [A*]